\documentclass[english,onecolumn]{IEEEtran}
\usepackage[T1]{fontenc}
\usepackage[latin9]{inputenc}
\usepackage{amsthm}
\usepackage{amsmath}
\usepackage{amssymb}
\usepackage{esint}
\PassOptionsToPackage{normalem}{ulem}
\usepackage{ulem}

\makeatletter
  \theoremstyle{plain}
  \newtheorem{lem}{\protect\lemmaname}
  \DeclareMathOperator{\sgn}{sgn}
\usepackage{cite}

\makeatother

\usepackage{babel}
\providecommand{\lemmaname}{Lemma}

\begin{document}

\title{Supporting Lemmas for RISE-based Control Methods}

\author{Rushikesh Kamalapurkar, Joel A. Rosenfeld, Justin Klotz, Ryan J.
Downey, and Warren E. Dixon}
\maketitle
\begin{abstract}
A class of continuous controllers termed Robust Integral of the Signum
of the Error (RISE) have been published over the last decade as a
means to yield asymptotic convergence of the tracking error for classes
of nonlinear systems that are subject to exogenous disturbances and/or
modeling uncertainties. The development of this class of controllers
relies on a property related to the integral of the signum of an error
signal. A proof for this property is not available in previous literature.
The stability of some RISE controllers is analyzed using differential
inclusions. Such results rely on the hypothesis that a set of points
is Lebesgue negligible. This paper states and proves two lemmas related
to the properties.
\end{abstract}

\section{Introduction}

A class of continuous controllers termed Robust Integral of the Signum
of the Error (RISE) have been published over the last decade as a
means to yield asymptotic convergence of the tracking error for classes
of nonlinear systems that are subject to exogenous disturbances and/or
modeling uncertainties. RISE-based controllers all exploit a property
that is instrumental for yielding an asymptotic result in the presence
of disturbances. Specifically, all RISE controllers exploit the fact
that the integral $\intop_{0}^{x}f^{\prime}\left(y\right)\sgn\left(f\left(y\right)\right)dy$
evaluates to $\left|f\left(x\right)\right|-\left|f\left(0\right)\right|$
as a means to prove the candidate Lyapunov function is positive definite
(cf. \cite{Xian2004,Xian2004c,Patre2008,Patre2008a,Patre2010,MacKunis2010,Patre2010a,MacKunis2010a,Patre2010b,Patre2011,Bhasin2011,Sharma2012,Shin.Kim.ea2012,Bhasin.Kamalapurkar.ea2013b}
and the references therein). However, no accessible proof of this
fact is available. Lemmas \ref{lem:absval} in this paper provides
a proof for the property. 

Motivated by robustness to measurement noise, the analysis of recent
RISE-based control designs is performed using non-smooth analysis
techniques (cf. \cite{Fischer.Kamalapurkar.ea2012b,Bialy.Andrews.ea2013,Bhasin.Kamalapurkar.ea2013b}).
To facilitate the non-smooth analysis, corollaries to the LaSalle-Yoshizawa
Theorem were recently published (cf. \cite{Fischer.Kamalapurkar.ea2013}).
The corollaries exploit the hypothesis that the generalized time derivative
of the Lyapunov function exists for almost all $t\in[0,\infty)$.
To satisfy the hypothesis, the fact that given a continuously differentiable
function $f:\left[0,\infty\right)\to\mathbb{R}$, then $\mu\left(\left\{ x\mid f\left(x\right)=0\wedge f^{\prime}\left(x\right)\neq0\right\} \right)=0$
where $\mu$ denotes the Lebesgue measure on $\left[0,\infty\right)$\sout{
}is used. Lemmas \ref{lem:MeasZero}-\ref{lem:MeasZeroGen} in this
paper provide proofs that validate the fact and further generalizations.
Throughout the paper, the notation $f^{\prime}$ is used to denote
the derivative of the function $f$ with respect to its argument,
and the notation $A^{c}$ is used to denote the complement of the
set $A$.

To facilitate Lyapunov-based stability analysis, a majority of RISE
controllers use the Mean Value Theorem to compute a strictly increasing
function that bounds the unknown functions in the system dynamics.
Lemma \ref{lem:rho} in this paper provides a constructive proof of
existence of a strictly increasing bound.

\section{Main Results}
\begin{lem}
\label{lem:absval}Let $f:\mathbb{R}_{+}\to\mathbb{R}$ be locally
absolutely continuous. Then, $\intop_{0}^{x}f^{\prime}\left(y\right)\sgn\left(f\left(y\right)\right)dy=\left|f\left(x\right)\right|-\left|f\left(0\right)\right|$.\end{lem}
\begin{IEEEproof}
Using the fundamental theorem of calculus, local absolute continuity
of $f$ implies that $f^{\prime}$ exists almost everywhere and that
$f^{\prime}$ is locally integrable. Since $\sgn\left(f\right)$ is
bounded, $f^{\prime}\sgn\left(f\right)$ is locally integrable. Thus,
for each $x$, $\intop_{0}^{x}f^{\prime}\left(y\right)\sgn\left(f\left(y\right)\right)dy<\infty$.
Since $f$ is continuous, $f^{-1}\left(\left\{ 0\right\} \right)$
is closed which means that $f\neq0$ only on an open subset $\mathcal{O}\subset\left[0,x\right]$.
The open subset $\mathcal{O}$ can be written as an at-most countable
union of mutually disjoint intervals. On some of these intervals $\sgn\left(f\right)=1$
and on the rest, $\sgn\left(f\right)=-1$. Define a sequence of functions
$\left(g_{n}\right)_{n=1}^{\infty}:\mathbb{R}_{+}\to\mathbb{R}$ as
\[
g_{n}\left(y\right)\triangleq\begin{cases}
\sum_{j=1}^{n}\mathbf{1}_{I_{j}}\left(y\right)-\sum_{k=1}^{n}\mathbf{1}_{I_{k}}\left(y\right) & if\: y\in\mathcal{O},\\
0 & otherwise.
\end{cases}
\]
where $I_{j}=\left(a_{j},b_{j}\right)$ and $I_{k}=\left(c_{k},d_{k}\right)$
are the (disjoint) intervals where $\sgn\left(f\right)$ is $+1$ or
$-1$, respectively, arranged such that $a_{j}>b_{j-1}$ for all $j>1$
and $c_{k}>d_{k-1}$ for all $k>1$, and $\mathbf{1}$ denotes the
indicator function defined as $\mathbf{1}_{I}\left(x\right)\triangleq\begin{cases}
1, & \text{if}\: x\in I\\
0 & \text{otherwise}
\end{cases}$. Then, $g_{n}\to \sgn\left(f\right)$ point-wise on $\left[0,x\right]$
as $n\to\infty$. Since $f^{\prime}$ is locally integrable and $\left[0,x\right]$
is compact, $f^{\prime}$ is integrable, and hence, essentially bounded
on $\left[0,x\right]$. Thus, $f^{\prime}g_{n}\to f^{\prime}\sgn\left(f\right)$
point-wise a.e. on $\left[0,x\right]$. Let $M=ess\sup_{y\in\left[0,x\right]}f^{\prime}\left(y\right)$.
Then, $\left|f^{\prime}\left(y\right)g_{n}\left(y\right)\right|\leq M$
for almost all $y\in\left[0,x\right]$, and hence, by the Dominated
convergence theorem \cite{Folland1999}, 
\begin{multline*}
\intop_{0}^{x}f^{\prime}\left(y\right)\sgn\left(f\left(y\right)\right)dy=\lim_{n\to\infty}\intop_{0}^{x}f^{\prime}\left(y\right)g_{n}\left(y\right)dy=\lim_{n\to\infty}\intop_{0}^{x}f^{\prime}\left(y\right)\left(\sum_{j=1}^{n}\mathbf{1}_{I_{j}}\left(y\right)-\sum_{k=1}^{n}\mathbf{1}_{I_{k}}\left(y\right)\right)dy\\
=\lim_{n\to\infty}\intop_{0}^{x}\left(\sum_{j=1}^{n}f^{\prime}\left(y\right)\mathbf{1}_{I_{j}}\left(y\right)-\sum_{k=1}^{n}f^{\prime}\left(y\right)\mathbf{1}_{I_{k}}\left(y\right)\right)dy=\lim_{n\to\infty}\biggl(\sum_{j=1}^{n}\intop_{0}^{x}f^{\prime}\left(y\right)\mathbf{1}_{I_{j}}\left(y\right)dy-\sum_{k=1}^{n}\intop_{0}^{x}f^{\prime}\left(y\right)\mathbf{1}_{I_{k}}\left(y\right)dy.\biggr).
\end{multline*}
Using the fundamental theorem of calculus, local absolute continuity
of $f$ implies that $\intop_{0}^{x}f^{\prime}\left(y\right)\mathbf{1}_{I_{j}}\left(y\right)dy=f\left(b_{j}\right)-f\left(a_{j}\right)$
and $\intop_{0}^{x}f^{\prime}\left(y\right)\mathbf{1}_{I_{k}}\left(y\right)dy=f\left(d_{k}\right)-f\left(c_{k}\right)$.
Thus
\[
\intop_{0}^{x}f^{\prime}\left(y\right)\sgn\left(f\left(y\right)\right)dy=\lim_{n\to\infty}\left(\sum_{j=1}^{n}\left(f\left(b_{j}\right)-f\left(a_{j}\right)\right)-\sum_{k=1}^{n}\left(f\left(d_{k}\right)-f\left(c_{k}\right)\right)\right).
\]
Since $f=0$ outside the open intervals $I_{j}$ and $I_{k}$, we
get $f\left(b_{1}\right)=f\left(d_{1}\right)=0$ and $f\left(a_{j}\right)=f\left(b_{j}\right)=f\left(c_{k}\right)=f\left(d_{k}\right)=0$
for all $2\leq j,k<\infty.$ Furthermore,
\begin{equation}
\lim_{j\to\infty}f\left(a_{j}\right)=\lim_{k\to\infty}f\left(c_{k}\right)=0,\label{eq:limzero}
\end{equation}
and
\begin{equation}
\intop_{0}^{x}f^{\prime}\left(y\right)\sgn\left(f\left(y\right)\right)dy=\underset{T1}{\underbrace{\lim_{j\to\infty}\left(f\left(b_{j}\right)-f\left(a_{j}\right)\right)-\lim_{k\to\infty}\left(f\left(d_{k}\right)-f\left(c_{k}\right)\right)}}-\underset{T2}{\underbrace{\left(f\left(a_{1}\right)-f\left(c_{1}\right)\right)}}.\label{eq:2}
\end{equation}
To evaluate $T2$, consider the following cases:

\textbf{Case 1: }$f\left(0\right)=0.$ In this case, since $f=0$
outside the open intervals $I_{j}$ and $I_{k}$, we get $f\left(a_{1}\right)=f\left(c_{1}\right)=0$. 

\textbf{Case 2: }$f\left(0\right)>0$. In this case, $a_{1}=0$, and
hence, $f\left(a_{1}\right)=f\left(0\right)$. Since $f=0$ outside
the open intervals $I_{k}$, $f\left(c_{1}\right)=0$. Thus, $f\left(0\right)>0\implies\left(f\left(a_{1}\right)-f\left(c_{1}\right)\right)=f\left(0\right).$

\textbf{Case 3: }$f\left(0\right)<0$. In this case, $c_{1}=0$, and
hence, $f\left(c_{1}\right)=f\left(0\right)$. Since $f=0$ outside
the open intervals $I_{j}$, $f\left(a_{1}\right)=0$. Thus, $f\left(0\right)<0\implies\left(f\left(a_{1}\right)-f\left(c_{1}\right)\right)=-f\left(0\right).$ 

Thus, \textbf{
\begin{equation}
f\left(a_{1}\right)-f\left(c_{1}\right)=\left|f\left(0\right)\right|.\label{eq:3}
\end{equation}
}

To evaluate $T1$, consider the following cases:

\textbf{Case 1: }$f\left(x\right)=0$. In this case, since $f=0$
outside the open intervals $I_{j}$ and $I_{k}$, we get $\lim_{j\to\infty}f\left(b_{j}\right)=\lim_{k\to\infty}f\left(d_{k}\right)=0,$
which from (\ref{eq:limzero}) implies $T1=0.$

\textbf{Case 2: }$f\left(x\right)>0$. In this case, $\lim_{j\to\infty}b_{j}=x$,
which from continuity of $f$ implies that $\lim_{j\to\infty}f\left(b_{j}\right)=f\left(x\right)$.
Furthermore, since $f=0$ outside the open intervals $I_{k}$, we
get $\lim_{k\to\infty}f\left(d_{k}\right)=0$. Thus, $T1=f\left(x\right).$

\textbf{Case 3: }$f\left(x\right)<0$. In this case, $\lim_{k\to\infty}d_{k}=x$,
which from continuity of $f$ implies that $\lim_{k\to\infty}f\left(d_{k}\right)=f\left(x\right)$.
Furthermore, since $f=0$ outside the open intervals $I_{j}$, we
get $\lim_{j\to\infty}f\left(b_{j}\right)=0$. Thus, $T1=-f\left(x\right).$

Thus,\textbf{
\begin{equation}
\lim_{j\to\infty}\left(f\left(b_{j}\right)-f\left(a_{j}\right)\right)-\lim_{k\to\infty}\left(f\left(d_{k}\right)-f\left(c_{k}\right)\right)=\left|f\left(x\right)\right|.\label{eq:4}
\end{equation}
}From \ref{eq:2}, \ref{eq:3}, and \ref{eq:4}, the required result
is follows.\end{IEEEproof}
\begin{lem}
\noindent \label{lem:MeasZero}Let $f:\left[0,\infty\right)\to\mathbb{R}$
be a continuously differentiable function. Then,
\begin{equation}
\mu\left(\left\{ x\mid f\left(x\right)=0\wedge f^{\prime}\left(x\right)\neq0\right\} \right)=0,\label{eq:conclusion}
\end{equation}
where $\mu$ denotes the Lebesgue measure on $\left[0,\infty\right)$.\end{lem}
\begin{IEEEproof}
Let $A\triangleq\left\{ x\mid f\left(x\right)=0\wedge f^{\prime}\left(x\right)\neq0\right\} \subseteq\left[0,\infty\right)$.
Note that $A=\left\{ f^{-1}\left(\left\{ 0\right\} \right)\right\} \cap\left\{ f^{\prime-1}\left(0\right)\right\} ^{c}$,
and hence, $A$ is measurable. The first step is to prove that all
the points in the set $A$ are isolated. That is,
\begin{equation}
\left(\forall a\in A\right)\left(\exists\epsilon>0\right)\mid\left(\left(\left(a-\epsilon,a+\epsilon\right)\cap A\right)\setminus\left\{ a\right\} =\phi\right).\label{eq:claim}
\end{equation}
The negation of (\ref{eq:claim}) is
\begin{equation}
\left(\exists a\in A\right)\mid\left(\forall\epsilon>0\right)\left(\left(\left(a-\epsilon,a+\epsilon\right)\cap A\right)\setminus\left\{ a\right\} \neq\phi\right).\label{eq:negclaim}
\end{equation}
For the sake of contradiction, assume that (\ref{eq:negclaim}) is
true. Thus, there exists a $b\in\left(\left(a-\epsilon,a+\epsilon\right)\cap A\right)\setminus\left\{ a\right\} $.
Without loss of generality, let $b>a$ and $f^{\prime}\left(a\right)>0.$
As $f$ is differentiable and $f\left(a\right)=f\left(b\right)=0$,
by the Mean Value Theorem, $\exists c\in\left(a,b\right)$ such that
\begin{equation}
f^{\prime}\left(c\right)=0.\label{eq:f'c=00003D0}
\end{equation}
By continuity of $f^{\prime}$ at $a,$ 
\[
\left(\forall\epsilon_{a}>0\right)\left(\exists\delta_{a}>0\right)\mid\left(\forall x\in\left[0,\infty\right)\right)\left(\left|x-a\right|<\delta_{a}\implies f^{\prime}\left(a\right)-\epsilon_{a}<f^{\prime}\left(x\right)<f^{\prime}\left(a\right)+\epsilon_{a}\right).
\]
In particular, pick $\epsilon_{a}=f^{\prime}\left(a\right).$ Then,
\begin{equation}
\left(\exists\delta_{a}>0\right)\mid\left(\forall x\in\left[0,\infty\right)\right)\left(\left|x-a\right|<\delta_{a}\implies f^{\prime}\left(x\right)>0\right).\label{eq:fprimebiggerthanzero}
\end{equation}
Now, pick $\epsilon=\mbox{\ensuremath{\delta}}_{a}$ in (\ref{eq:negclaim}).
Thus, $b\in\left(\left(a-\delta_{a},a+\delta_{a}\right)\cap A\right)\setminus\left\{ a\right\} $.
Since $\left|b-a\right|<\delta_{a}$, and $c\in\left(a,b\right)$,
it can be concluded that $\left|c-a\right|<\delta_{a}$. Thus, from
(\ref{eq:fprimebiggerthanzero}), $f^{\prime}\left(c\right)>0$, which
contradicts (\ref{eq:f'c=00003D0}). Thus, all the points in the set
$A$ are isolated, and hence, $A$ is a discrete set. Using the fact
that any discrete subset of $\mathbb{R}$ is countable, (\ref{eq:conclusion})
follows.
\end{IEEEproof}
The following two lemmas generalize the above result.
\begin{lem}
\label{lem:MeasZero1}Let $f:\mathbb{R}\to\mathbb{R}$ be an everywhere
differentiable function. The set $E=\{a\in\mathbb{R}:f(a)=0\ and\ f^{\prime}(a)\neq0\}$
is countable.\end{lem}
\begin{IEEEproof}
If $E$ is empty, then it is countable. Now suppose that $E$ is nonempty.
We will show that $E$ is composed of only isolated points. Let $a\in E$,
and consider the first order Taylor expansion of $f$: 
\[
f(x)=f(a)+f^{\prime}(a)(x-a)+\epsilon(x)
\]

First note that:

\[
\frac{\epsilon(x)}{(x-a)}=\frac{f(x)-f(a)-f^{\prime}(a)(x-a)}{(x-a)}=\frac{f(x)-f(a)}{x-a}-f^{\prime}(a)\to0
\]
as $x\to a$.

Now pick a $\delta>0$ such that $|\epsilon(x)/(x-a)|<|f^{\prime}(a)|$
for $x\in(a-\delta,a+\delta)$. For this neighborhood we have (with
$x\neq a$): 
\[
|f(x)|=\left|f^{\prime}(a)(x-a)+\frac{\epsilon(x)}{x-a}(x-a)\right|\ge|x-a|\left|\left|f^{\prime}(a)\right|-\left|\frac{\epsilon(x)}{x-a}\right|\ \right|>0.
\]

Therefore we have $f(x)\neq0$ in the neighborhood $(a-\delta,a+\delta)$
unless $x=a$. Hence each point in $E$ is isolated, and therefore
$E$ is countable. By the proof of this theorem we can also weaken
the everywhere differentiability and find that the set: 
\[
E=\{a\in\mathbb{R}:f\text{ is differentiable at }a,f(a)=0,f^{\prime}(a)\neq0\}
\]
is countable.\end{IEEEproof}
\begin{lem}
\label{lem:MeasZeroGen}Let $f:\mathbb{R}\to\mathbb{R}$ be a function.
Consider the set 
\[
E=\left\{ a\in\mathbb{R}:\lim\inf_{x\to a}\frac{f(x)-f(a)}{x-a}>0\text{ or }\lim\sup_{x\to a}\frac{f(x)-f(a)}{x-a}<0,f(a)=0\right\} .
\]
This set is countable.\end{lem}
\begin{IEEEproof}
Suppose that $E$ has some accumulation point $a\in\mathbb{R}$. This
means there is a sequence of points $\{a_{n}\}\subset E$ such that
$\lim a_{n}=a$. Without loss of generality we may assume that 
\[
\lim\inf_{n\to a}\frac{f(x)-f(a)}{x-a}>0.
\]
This means for any sequence $x_{n}\to a$ for which the sequence $\frac{f(x_{n})-f(a)}{x_{n}-a}$
converges, the limit of that convergent sequence is greater than zero.

However, since $f(a_{n})=0$ and $f(a)=0$ we have 
\[
\frac{f(a_{n})-f(a)}{a_{n}-a}=0
\]
for all $n$. A contradiction. Thus every point is isolated and $E$
is countable.\end{IEEEproof}
\begin{lem}
\label{lem:rho}If $B_{r}\subset \mathbb{R}^n$ denotes the closed ball of radius $r>0$ centered at the origin, and $f:\mathbb{R}^n\to\mathbb{R}^{m}=\left[f_{1},f_{2},\cdots,f_{m}\right]^{T}$
is continuously differentiable for all $i = 1,\ldots,m$, then there exists a continuous strictly increasing function $\rho:[0,\infty)\to[0,\infty)$ such that $\left\Vert f\left(x\right)-f\left(x_{d}\right)\right\Vert \leq\rho\left(\left\Vert x-x_{d}\right\Vert \right)\left\Vert x-x_{d}\right\Vert $
for all $x\in \mathbb{R}^n$ and $x_{d}\in B_{r}$.\end{lem}
\begin{IEEEproof}
Using the Mean Value Theorem, $\forall i=1,\cdots,m,$ and for all
$x,x_{d}\in \mathbb{R}^n$ there exist $0<c_{i}<1$ such that 
\[
f_{i}\left(x\right)-f_{i}\left(x_{d}\right)=\left.\nabla f_{i}\right|_{c_{i}x+\left(1-c_{i}\right)x_{d}}\cdot\left(x-x_{d}\right).
\]
Using the Cauchy-Schwarz inequality, 
\begin{align*}
\left\Vert f\left(x\right)-f\left(x_{d}\right)\right\Vert  & =\sqrt{\sum_{i=1}^{m}\left|f_{i}\left(x\right)-f_{i}\left(x_{d}\right)\right|^{2}},\\
 & =\sqrt{\sum_{i=1}^{m}\left(\left.\nabla f_{i}\right|_{c_{i}x+\left(1-c_{i}\right)x_{d}}\cdot\left(x-x_{d}\right)\right)^{2}},\\
 & \leq\sqrt{\sum_{i=1}^{m}\left(\left\Vert \left.\nabla f_{i}\right|_{c_{i}\left(x-x_{d}\right)+x_{d}}\right\Vert ^{2}\right)}\left\Vert \left(x-x_{d}\right)\right\Vert ,\\
 & \leq\sqrt{\sum_{i=1}^{m}\max_{i}\left\Vert \left.\nabla f_{i}\right|_{c_{i}\left(x-x_{d}\right)+x_{d}}\right\Vert ^{2}}\left\Vert \left(x-x_{d}\right)\right\Vert .\\
 & =G_{1}\left(x-x_{d},x_{d}\right)\left\Vert \left(x-x_{d}\right)\right\Vert 
\end{align*}
where the function $G_{1}:\mbox{\ensuremath{\mathbb{R}}}^{n}\times \mathbb{R}^n\to[0,\infty)$
is defined as $G_{1}\left(x-x_{d},x_{d}\right)=\sqrt{m\left(\max_{i}\left\Vert \left.\nabla f_{i}\right|_{c_{i}\left(x-x_{d}\right)+x_{d}}\right\Vert ^{2}\right)}$.

For a given $x$ and $x_d$, define a set $B\subseteq\mathbb{R}^n\times\mathbb{R}^n$
as $$B\triangleq\left\{ \left(\sigma,\omega\right)\in\mathbb{R}^n\times\mathbb{R}^n\mid 0 \leq \|\sigma\|\leq\left\Vert x-x_{d}\right\Vert ,0\leq\|\omega\|\leq\left\Vert x_{d}\right\Vert \right\}.$$
Since $f$ is continuously differentiable, $\nabla f_{i}$ is continuous for all $i=1,\cdots,m$. Thus, $G_{1}$ is continuous on the compact
set $B$. Therefore, $G_{1}$ attains its maximum value on the set $B$. Define the function $G_{2}:\mathbb{R}_{+}\times\mathbb{R}_{+}\to[0,\infty)$
as $$G_{2}\left(\left\Vert x-x_{d}\right\Vert ,\left\Vert x_{d}\right\Vert \right)=\max_{\left(\sigma,\omega\right)\in B}G_{1}\left(\sigma,\omega\right).$$ Note that $G_{2}$ is nondecreasing and continuous in both its arguments. Furthermore, since $\left\Vert x_{d}\right\Vert \leq r$
for all $x_{d}\in B_{r}$, 
\begin{equation}
  G_{2}\left(\left\Vert x-x_{d}\right\Vert ,r\right)\geq G_{2}\left(\left\Vert x-x_{d}\right\Vert ,\left\Vert x_{d}\right\Vert \right)\label{eq:Bound}
\end{equation}
for all $x\in \mathbb{R}^n$, $x_{d}\in B_{r}$. A strictly increasing continuous function $\rho:[0,\infty)\to[0,\infty)$
can now be defined as $$\rho\left(\left\Vert x-x_{d}\right\Vert \right)\triangleq G_{2}\left(\left\Vert x-x_{d}\right\Vert ,r\right)+\left\Vert x-x_{d}\right\Vert. $$
Using (\ref{eq:Bound}) and the definitions of $G_1$ and $G_2$, 
\[
\left\Vert f\left(x\right)-f\left(x_{d}\right)\right\Vert \leq\rho\left(\left\Vert x-x_{d}\right\Vert \right)\left\Vert x-x_{d}\right\Vert ,
\]
for all $x\in \mathbb{R}^n$ and $x_{d}\in B_{r}$.
\end{IEEEproof}

\section*{Acknowledgments}

The authors thank Max L. Gardenswartz, Dr. Michael Jury, Dr. Marcio de Queiroz and Dr.
Xiaoyu Cai for their valuable inputs.

\bibliographystyle{IEEEtran}
\bibliography{master,ncr}

\begin{thebibliography}{10}
\def\url#1{}
\csname url@samestyle\endcsname
\providecommand{\newblock}{\relax}
\providecommand{\bibinfo}[2]{#2}
\providecommand{\BIBentrySTDinterwordspacing}{\spaceskip=0pt\relax}
\providecommand{\BIBentryALTinterwordstretchfactor}{4}
\providecommand{\BIBentryALTinterwordspacing}{\spaceskip=\fontdimen2\font plus
\BIBentryALTinterwordstretchfactor\fontdimen3\font minus
  \fontdimen4\font\relax}
\providecommand{\BIBforeignlanguage}[2]{{%
\expandafter\ifx\csname l@#1\endcsname\relax
\typeout{** WARNING: IEEEtran.bst: No hyphenation pattern has been}%
\typeout{** loaded for the language `#1'. Using the pattern for}%
\typeout{** the default language instead.}%
\else
\language=\csname l@#1\endcsname
\fi
#2}}
\providecommand{\BIBdecl}{\relax}
\BIBdecl

\bibitem{Xian2004}
B.~Xian, D.~M. Dawson, M.~S. de~Queiroz, and J.~Chen, ``A continuous asymptotic
  tracking control strategy for uncertain nonlinear systems,'' \emph{IEEE
  Trans. Autom. Control}, vol.~49, pp. 1206--1211, 2004.

\bibitem{Xian2004c}
B.~Xian, M.~S. de~Queiroz, D.~M. Dawson, and M.~McIntyre, ``A discontinuous
  output feedback controller and velocity observer for nonlinear mechanical
  systems,'' \emph{Automatica}, vol.~40, no.~4, pp. 695--700, 2004.

\bibitem{Patre2008}
P.~M. Patre, W.~MacKunis, K.~Kaiser, and W.~E. Dixon, ``Asymptotic tracking for
  uncertain dynamic systems via a multilayer neural network feedforward and
  {RISE} feedback control structure,'' \emph{IEEE Trans. Automat. Control},
  vol.~53, no.~9, pp. 2180--2185, 2008.

\bibitem{Patre2008a}
P.~M. Patre, W.~Mackunis, C.~Makkar, and W.~E. Dixon, ``Asymptotic tracking for
  systems with structured and unstructured uncertainties,'' \emph{IEEE Trans.
  Control Syst. Technol.}, vol.~16, pp. 373--379, 2008.

\bibitem{Patre2010}
P.~M. Patre, W.~MacKunis, K.~Dupree, and W.~E. Dixon, \emph{{RISE}-Based Robust
  and Adaptive Control of Nonlinear Systems}.\hskip 1em plus 0.5em minus
  0.4em\relax Birkhäuser: Boston, 2010.

\bibitem{MacKunis2010}
\BIBentryALTinterwordspacing
W.~MacKunis, K.~Kaiser, Z.~D. Wilcox, and W.~E. Dixon, ``Global adaptive output
  feedback tracking control of an unmanned aerial vehicle,'' \emph{IEEE Trans.
  Control Syst. Technol.}, vol.~18, no.~6, pp. 1390--1397, 2010.
  \url{http://ncr.mae.ufl.edu/papers/CST10-1.pdf}
\BIBentrySTDinterwordspacing

\bibitem{Patre2010a}
\BIBentryALTinterwordspacing
P.~Patre, S.~Bhasin, Z.~D. Wilcox, and W.~E. Dixon, ``Composite adaptation for
  neural network-based controllers,'' \emph{IEEE Trans. Automat. Control},
  vol.~55, no.~4, pp. 944--950, 2010.
  \url{http://ncr.mae.ufl.edu/papers/tac10.pdf}
\BIBentrySTDinterwordspacing

\bibitem{MacKunis2010a}
\BIBentryALTinterwordspacing
W.~MacKunis, P.~Patre, M.~Kaiser, and W.~E. Dixon, ``Asymptotic tracking for
  aircraft via robust and adaptive dynamic inversion methods,'' \emph{IEEE
  Trans. Control Syst. Technol.}, vol.~18, no.~6, pp. 1448--1456, 2010.
  \url{http://ncr.mae.ufl.edu/papers/CST10-2.pdf}
\BIBentrySTDinterwordspacing

\bibitem{Patre2010b}
\BIBentryALTinterwordspacing
P.~Patre, W.~Mackunis, M.~Johnson, and W.~E. Dixon, ``Composite adaptive
  control for {E}uler-{L}agrange systems with additive disturbances,''
  \emph{Automatica}, vol.~46, no.~1, pp. 140--147, 2010.
  \url{http://ncr.mae.ufl.edu/papers/auto10.pdf}
\BIBentrySTDinterwordspacing

\bibitem{Patre2011}
\BIBentryALTinterwordspacing
P.~Patre, W.~Mackunis, K.~Dupree, and W.~E. Dixon, ``Modular adaptive control
  of uncertain {E}uler-{L}agrange systems with additive disturbances,''
  \emph{IEEE Trans. Automat. Control}, vol.~56, no.~1, pp. 155--160, 2011.
  \url{http://ncr.mae.ufl.edu/papers/tac11.pdf}
\BIBentrySTDinterwordspacing

\bibitem{Bhasin2011}
S.~Bhasin, N.~Sharma, P.~Patre, and W.~E. Dixon, ``Asymptotic tracking by a
  reinforcement learning-based adaptive critic controller,'' \emph{J. Control
  Theory Appl.}, vol.~9, no.~3, pp. 400--409, 2011.

\bibitem{Sharma2012}
N.~Sharma, S.~Bhasin, Q.~Wang, and W.~E. Dixon, ``{RISE}-based adaptive control
  of a control affine uncertain nonlinear system with unknown state delays,''
  \emph{IEEE Trans. Automat. Control}, vol.~57, no.~1, pp. 255--259, Jan. 2012.

\bibitem{Shin.Kim.ea2012}
J.~Shin, H.~Kim, Y.~Kim, and W.~E. Dixon, ``Autonomous flight of the
  rotorcraft-based {UAV} using {RISE} feedback and {NN} feedforward terms,''
  \emph{IEEE Trans. Control Syst. Technol.}, vol.~20, no.~5, pp. 1392--1399,
  2012.

\bibitem{Bhasin.Kamalapurkar.ea2013b}
S.~Bhasin, R.~Kamalapurkar, H.~T. Dinh, and W.~Dixon, ``Robust
  identification-based state derivative estimation for nonlinear systems,''
  \emph{IEEE Trans. Automat. Control}, vol.~58, no.~1, pp. 187--192, 2013.

\bibitem{Fischer.Kamalapurkar.ea2012b}
N.~Fischer, R.~Kamalapurkar, N.~Sharma, and W.~E. Dixon, ``Rise-based control
  of an uncertain nonlinear system with time-varying state delays,'' in
  \emph{Proc. IEEE Conf. Decis. Control}, Maui, HI, Dec. 2012, pp. 3502--3507.

\bibitem{Bialy.Andrews.ea2013}
B.~Bialy, L.~Andrews, J.~Curtis, and W.~E. Dixon, ``Saturated rise tracking
  control of store-induced limit cycle oscillations,'' in \emph{Proc. AIAA
  Guid., Navig., Control Conf., AIAA 2013-4529}, August 2013.

\bibitem{Fischer.Kamalapurkar.ea2013}
N.~Fischer, R.~Kamalapurkar, and W.~E. Dixon, ``La{S}alle-{Y}oshizawa
  corollaries for nonsmooth systems,'' \emph{IEEE Trans. Automat. Control},
  vol.~58, no.~9, pp. 2333--2338, 2013.

\bibitem{Folland1999}
G.~B. Folland, \emph{Real analysis: modern techniques and their applications},
  2nd~ed., ser. Pure and applied mathematics.\hskip 1em plus 0.5em minus
  0.4em\relax Wiley, 1999.

\end{thebibliography}

\end{document}